\documentclass[apj]{emulateapjmod}
\slugcomment{Preprint: accepted by ApJ, August 28, 2006}

\journalinfo{\scriptsize \copyright\,2006. The American Astronomical Society. All rights reserved.}

\shorttitle{Dust and PAHs in the SNR N132D}

\shortauthors{Tappe, Rho, \& Reach}

\begin{document}

\newcommand{\mic}{\ensuremath{{\, \rm \mu m}}}
\newcommand{\mj}{\ensuremath{{\rm\,MJy\!/\!sr^{-1}}}}
\newcommand{\wms}{\ensuremath{{\rm\,W\,m^{-2}sr^{-1}}}}
\newcommand{\km}{\ensuremath{{\rm\,km\,s^{-1}}}}
\newcommand{\ccm}{\ensuremath{{\rm\,cm^{-3}}}}
\newcommand{\scm}{\ensuremath{{\rm\,cm^{-2}}}}
\newcommand{\hii}{\ion{H}{2}}
\newcommand{\htwo}{\ensuremath{\rm H_2}}
\newcommand{\oiiif}{\ion{[O}{3}]}
\newcommand{\siif}{\ion{[S}{2}]}
\newcommand{\siiif}{\ion{[S}{3}]}
\newcommand{\siliif}{\ion{[Si}{2}]}
\newcommand{\hi}{\ion{H}{1}}
\newcommand{\feiif}{\ion{[Fe}{2}]}
\newcommand{\feiiif}{\ion{[Fe}{3}]}
\newcommand{\fevif}{\ion{[Fe}{6}]}
\newcommand{\neiiif}{\ion{[Ne}{3}]}
\newcommand{\oivf}{\ion{[O}{4}]}
\newcommand{\msun}{\ensuremath{{\, \rm M_\odot}}}

\defcitealias{vanke2000}{VHP}

\title{Shock processing of interstellar dust and polycyclic aromatic hydrocarbons in the supernova remnant N132D}
\author{A.~Tappe, J.~Rho, W.~T.~Reach}
\affil{{\it Spitzer} Science Center, California Institute of Technology, Pasadena, CA\,91125}
\email{tappe@ipac.caltech.edu}

\begin{abstract}
We observed the oxygen-rich Large Magellanic Cloud (LMC) supernova remnant N132D (SNR\,0525--69.6), using all
instruments onboard the {\it Spitzer Space Telescope}, IRS, IRAC, and MIPS (Infrared Spectrograph, Infrared Array
Camera, Multiband Imaging Photometer for Spitzer). The 5--40\mic\ IRS spectra toward the southeastern shell of the
remnant show a steeply rising continuum with \neiiif\ and \oivf\ as well as PAH emission. We also present the
spectrum of a fast moving ejecta knot, previously detected at optical wavelengths, which is dominated by strong
\neiiif\ and \oivf\ emission lines. We interpret the continuum as thermal emission from swept-up, shock-heated
dust grains in the expanding shell of N132D, which is clearly visible in the MIPS 24\mic\ image. A
\mbox{15--20\mic} emission hump appears superposed on the dust continuum, and we attribute this to PAH
\mbox{C-C-C} bending modes. We also detect the well-known 11.3\mic\ PAH C-H bending feature, and find the
integrated strength of the 15--20\mic\ hump about a factor of seven stronger than the 11.3\mic\ band in the shell
of the remnant. IRAC 3--9\mic\ images do not show clear evidence of large-scale, shell-like emission from the
remnant, partly due to confusion with the ambient ISM material. However, we identified several knots of shocked
interstellar gas based on their distinct infrared colors. We discuss the bright infrared continuum and the
polycyclic aromatic hydrocarbon features with respect to dust processing in young supernova remnants.

\end{abstract}

\keywords{infrared: ISM --- supernova remnants --- supernovae: individual (N132D)}

\section{Introduction}
\label{sec;introduction}
Supernovae influence the chemistry and physics in the interstellar medium (ISM) from galactic scales down to the
atomic level. Primarily, it is the energy input of supernova explosions that drives important processes in the
ISM: strong shock waves accelerate, heat, and compress interstellar gas, destroy dust grains, accelerate cosmic
rays, and affect interstellar chemistry \citep[cf.][]{draine1993}.

The role that supernovae play in the formation, destruction, and processing of polycyclic aromatic hydrocarbons
(PAHs) is largely unexplored. PAHs are believed to be the carrier of the unidentified infrared (UIR) bands,
well-studied emission features near 3.3, 6.2, 7.7, 8.6, and 11.3\mic\ attributed to the C-C and C-H stretching and
bending modes of these molecules \citep{allam1989,puget1989}. It is established that interstellar grains are
readily destroyed by strong shock waves \citep[e.g.][]{tiele1994}, and it was suggested that grain destruction
through shocks may be a source of PAH molecules \citep{jones1996}. Although PAH features have been observed in a
wide variety of sources \citep[cf.][]{vandi2004}, no PAH emission has yet been detected in SNRs as far as we are
aware, however, \citet{reach2006} identified four SNRs whose infrared colors suggest emission from PAHs.

More recently, spectral features at 15--20\mic\ including  broad plateaus and distinct, narrower features, most
notably near 16.4 and 17.4\mic, were detected in conjunction with the well-known PAH features \citep[][and see
\citealp{peete2004} for a summary]{mouto2000,vanke2000,armus2004,morri2004,smith2004,werne2004}. The 15--20\mic\
features are generally attributed to \mbox{C-C-C} bending modes of large PAH molecules
\citep[cf.][]{allam1989,vanke2000,peete2004}. The relatively narrow observed widths of the discrete 16.4 and
17.4\mic\ bands suggest that they arise from individual PAH molecules in the gas phase \citep{peete2004}. There is
also evidence that the appearance, relative strengths, and profiles of the PAH features are coupled to the local
physical conditions \citep{hony2001,peete2002,vandi2004,werne2004}.

In this paper, we report the detection of weak PAH features near 6.2 and 7.7--8.6\mic, at 11.3\mic, and a
prominent, broad hump at 15--20\mic\ from the young supernova remnant N132D (SNR\,0525$-$69.6) in the Large
Magellanic Cloud (LMC). The age of N132D determined from the ejecta kinematics is $\sim$\,2500\,yr
\citep{morse1995}, and its spatial extent is about 100\arcsec, or 25\,pc assuming a LMC distance of 50\,kpc. N132D
belongs to the small class of young, oxygen-rich SNRs that are presumably the product of core-collapse supernovae,
with the most famous Galactic example being Cas\,A. For the progenitor of N132D, \citet{blair2000} suggested a W/O
star of 30--35\msun\ based on relative elemental abundances and comparisons to stellar nucleosynthesis models,
which may have exploded as a Type Ib supernova. N132D shows a radio spectral index of $-0.7$ \citep{dicke1995},
consistent with its overall shell-like morphology. It is one of the brightest X-ray sources in the LMC
\citep{hughe1987,hwang1993,favat1997,hughe1998}, and has also been investigated in great detail in the UV/optical
\citep{blair1994,morse1995,suthe1995,morse1996,blair2000}. The large-scale morphology of the gas in N132D shows
three types of line-emitting regions in the optical \citep{morse1996}: (1) red- and blueshifted oxygen-rich ejecta
filaments in the center, with a total velocity range $\sim$\,4400\km\ \citep{morse1995}, (2) shocked interstellar
clouds within the remnant perimeter and in the outer rim, and (3) a photoionized precursor around the outer edge
of the remnant.

We observed N132D in the infrared with the {\it Spitzer Space Telescope} \citep{werne2004i}. In
Section~\ref{sec;observations}, we describe our observations and the data reduction procedure. We took particular
care of the background subtraction in order to avoid confusion with foreground and background emission (Section
\ref{sec;background}). We present our results from the imaging and the spectroscopy in Section \ref{sec;results},
and give a detailed discussion of the IR emission features and a comparison to PAH features observed in other
astronomical objects in Section~\ref{sec;discussion}.

\section{Observations and data analysis}
\label{sec;observations}
We observed N132D with all instruments onboard the {\it Spitzer Space Telescope}: IRS \citep[Infrared
Spectrograph,][]{houck2004} 2004 December~13, IRAC \citep[Infrared Array Camera,][]{fazio2004} 2004 November~28,
and MIPS \citep[Multiband Imaging Photometer for Spitzer,][]{rieke2004} 2004 November~7 ({\it Spitzer} aorkeys
11019776, 11018496, and 11022336, respectively). We used the IRS in its low resolution setting, yielding a
wavelength coverage of 5.2--8.7/7.4--14.5\mic\ with the short-low (SL2/SL1) and 14.0--21.3/19.5--38.0\mic\ with
the long-low (LL2/LL1) modules. The nominal spectral resolution is $R=64$--128 as a function of wavelength for
each module and order, e.g.~$R\sim100$ at 18\mic. The spatial sampling of the IRS is 5.1/1.8\arcsec\ per pixel in
the LL/SL modules, and the spatial resolution is diffraction limited by the 85\,cm primary mirror. The four IRAC
bands cover the wavelength ranges 3.2--4.0, 4.0--5.0, 5.0--6.4, and 6.4--9.4\mic, respectively, with a pixel size
of 1.2\arcsec. The three MIPS detector bands cover 20.8--26.1, 61--80, and 140--174\mic, with respective pixel
sizes of 2.6, 5.2, and 17\arcsec\ in our configuration. The approximate total exposure times are 270\,s for each
IRAC band, 48, 38, and 25\,s for the MIPS bands, 280\,s for each IRS SL, and 360\,s for each LL module.

\subsection{General data reduction procedure}
\label{sec;datareduction}
The pipeline processed MIPS and IRAC image mosaics were of good quality, and we used them without further
processing. The analysis of the IRS spectra is complicated by the fact that N132D is a non-uniform, extended
source superposed with a complex background adjacent to a molecular cloud (see Fig.~\ref{fig:n132d}). Moreover,
the {\it Spitzer} science pipeline is optimized for point sources, and the extraction of extended sources from the
default pipeline products results in biased spectra. Our principal data reduction strategy for the IRS spectra
consists therefore of the following steps:

\begin{figure}[!tbp]
\epsscale{1.15}
\centering
 \plotone{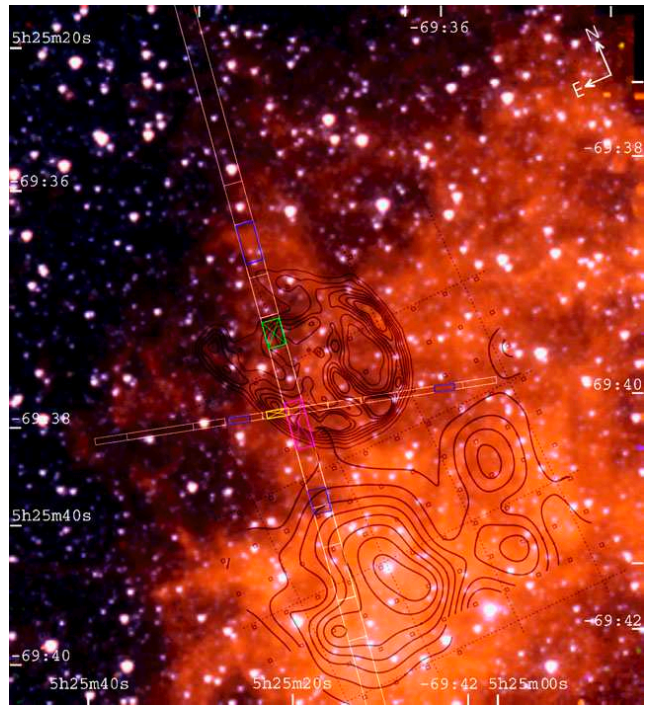} \caption{N132D: {\it Spitzer} IRAC 3--9\mic\ color composite image (3.6\mic=blue,
4.5\mic=green, 5.7\mic=orange, and 8.0\mic=red); contours at the image center are X-ray {\it ROSAT} counts of
N132D, and contours south of the remnant are CO $J=2\!\rightarrow\!1$ emission \citep[both adapted
from][]{banas1997}. The IRS slit coverage is outlined with polygons: SL=west--east, LL=north--south; yellow (SL)
and magenta (LL) boxes indicate the extracted parts for the southeastern rim spectrum, the green box marks the
position of an ejecta knot detected in the IRS LL spectra, and blue boxes mark the extracted positions for the
local background (Sect.~\ref{sec;background}).}
 \label{fig:n132d}
\end{figure}

(1) starting from the S13.2.0 pipeline products, we interpolated bad pixels flagged by the pipeline, and removed
rogue pixels by subtracting nearby sky positions taken from the {\it Spitzer} archive for each SL/LL module (see
Sect.~\ref{sec;background} for details). Any remaining rogues were removed with the IRSCLEAN\footnote{available
through the Spitzer Science Center at Caltech, see {\tt http://ssc.spitzer.caltech.edu/archanaly/contributed/};
IRSCLEAN is developed by Jim Ingalls with contributions from the Cornell IRS instrument team.} tool. Rogue pixels
are single detector pixels that show time-variable and abnormally high flux values.

(2) we extracted the spectra with SPICE\footnote{Spitzer IRS Custom Extraction software available through the
Spitzer Science Center at Caltech, see {\tt http://ssc.spitzer.caltech.edu/postbcd/spice.html}}\,v.1.3beta1 using
the default extraction widths, and imported the spectra into SMART\footnote{The Spectroscopy Modelling Analysis
and Reduction Tool was developed by the IRS Team at Cornell University, and is available through the Spitzer
Science Center at Caltech.} \citep{higdo2004} for postprocessing;

(3) subtraction of a local background spectrum (see Sect.~\ref{sec;background});

(4) basic aperture and slit loss correction assuming a uniform, extended source, to account for the bias
introduced by the point source optimized {\it Spitzer} science pipeline. We derived a correction function by
modelling the {\it Spitzer} point-spread-function (PSF) for a range of wavelengths with the software STINYTIM$^1$
developed by John Krist, and then dividing the fraction of the PSF admitted by the slit as defined by the
slit-width and wavelength dependent extraction aperture by the sum over the whole PSF. The resulting flux
correction functions have wavelength dependent values of 0.8--0.6 for SL and 0.85--0.65 for LL. Further division
of the corrected flux density by the extracted area given by the slit-width and the extraction aperture yields
spectra in surface brightness units, i.e.~flux density per steradian, which can be directly compared to the IRAC
and MIPS fluxes. Note that IRAC fluxes are calibrated for point sources, but established point-to-extended source
flux correction factors are available \citep{reach2005};

(5) we removed fringes that appear in LL1 at 19.5--24.0\mic\ with the SMART defringing tool, averaged multiple
spectra of the same region, and finally merged SL and LL spectra. Note that we extracted the SL and LL spectra of
the southeastern rim from two slightly offset positions, because the small overlap region covered by both slits is
positioned just outside the brightest regions in the southeastern rim (cf.~Fig.~\ref{fig:n132d}). The two
extracted regions are separated by about 10\arcsec, and the MIPS\,24\mic\ fluxes in both regions differ by only 10
percent, whereas the IRAC channel 3 and 4 fluxes are comparable.

\subsection{Background removal}
\label{sec;background}
A meaningful background subtraction of the spectra is one of the most critical and difficult tasks, and of major
importance for this work. Commonly, either a spatially close spectral nod or a dedicated off-source position are
subtracted to remove the background. Both of these methods are problematic in our case, since we are facing a
complex, non-uniform background (cf.~Fig.~\ref{fig:n132d}).

As a first step, we subtracted archival off-source sky exposures from all spectral modules
(cf.~Sect.~\ref{sec;datareduction}, item 1). The selected sky positions are devoid of continuum and line emission,
and were observed close in time and position with respect to our observations (\raisebox{0.5mm}{\scriptsize
${<}$}\,3\,hrs, $\sim$\,10\arcmin). The sky position {\it Spitzer} aorkeys are 10964224, 12935424, and 12936960
for SL2, 12936960 for SL1, and 12935424 for LL2/LL1. The sky subtraction removes rogue pixels, zodiacal light
emission, and flat-field imperfections, which are particularly noticeable in the SL modules.

In a subsequent step, we subtracted a rogue-cleaned local background spectrum from the extracted source spectra
for each spectral module (cf.~Sect.~\ref{sec;datareduction}, item 3). We selected positions that lie close enough
to the source to represent the local background but sufficiently far from the rim to avoid the influence of the
radiative precursor that extends noticeably beyond the edge of the remnant (see Sect.~\ref{sec;IRAC}). The
background positions are depicted as blue boxes in Fig.~\ref{fig:n132d}. We averaged two background positions from
opposite sides of N132D for SL2 and LL2/LL1, while ensuring that both positions show similar spectra. For SL1, we
only have slit coverage east of the remnant, and we adopted the same background position as for SL2. The overall
background spectrum extracted from these positions is shown in Fig.~\ref{fig:n132dspec+backgr}. We estimate a 30
percent uncertainty in this background spectrum due to spatial variations by using slightly different extraction
positions (approximately $\pm 10$\arcsec; blue boxes in Fig.~\ref{fig:n132d}) and two background positions on
opposite sides of the remnant.

\subsection{Analysis of emission lines}
\label{sec:analysis_emissionlines}
We measured emission line fluxes, central wavelengths, and full-widths at half-maximum (FWHM) from line fitting
with the IRAF `spectool' package\footnote{IRAF (Image Reduction and Analysis Facility) and `spectool' are
maintained by the IRAF programming group at the National Optical Astronomy Observatory (NOAO) in Tucson,
Arizona.}. Due to the low resolution of the IRS spectrograph in our setting, $R\sim100$ at 18\mic, the lines are
usually unresolved or marginally resolved, and represent a convolution of the true line profile with the
instrumental profile. We used Gaussians to fit the line profiles of atomic lines/\htwo\ and Lorentzians for
discrete PAH features. Our quoted FWHM are not deconvolved from the instrumental IRS profile.

We also extracted spectra from different spatial positions at $\approx25$\arcsec\ intervals along the LL
spectrograph slit in order to study the spatial variation of line fluxes across the supernova remnant and the
molecular cloud to the south (see Sect.~\ref{sec:discussion_spatialvariation},
Fig.~\ref{fig:n132dspec_spatialvariation}). For this particular task, we employed a slightly different data
extraction strategy. In order to improve the spatial resolution, we narrowed the default SPICE extraction aperture
by a factor of two, giving extraction widths of about 12\arcsec\ for LL2 and 20\arcsec\ for LL1 near the central
wavelengths of each module. This procedure systematically reduces measured fluxes by about a factor of two
compared to the calibrated default extractions, which is of no concern as long as we compare only relative fluxes
of extended emission extracted in that same fashion. We subtracted the continuum by fitting a low-order Legendre
polynomial using spectral regions that are free of emission lines, usually below 15 and above 19\mic. For the
southeastern rim of N132D, this method gives a similar continuum fit for the LL2 14--21\mic\ region than the
modified-blackbody fit shown in Fig.~\ref{fig:n132dspec+bbfit}. We did {\it not} apply a local background
subtraction here, which means that we compare emission line fluxes including fore- and background for each spatial
position along the LL slit.


\section{Results}
\label{sec;results}
\subsection{IRAC and MIPS imaging}
\label{sec;IRAC}
\begin{figure*}[!tbp]
\centering
 \plotone{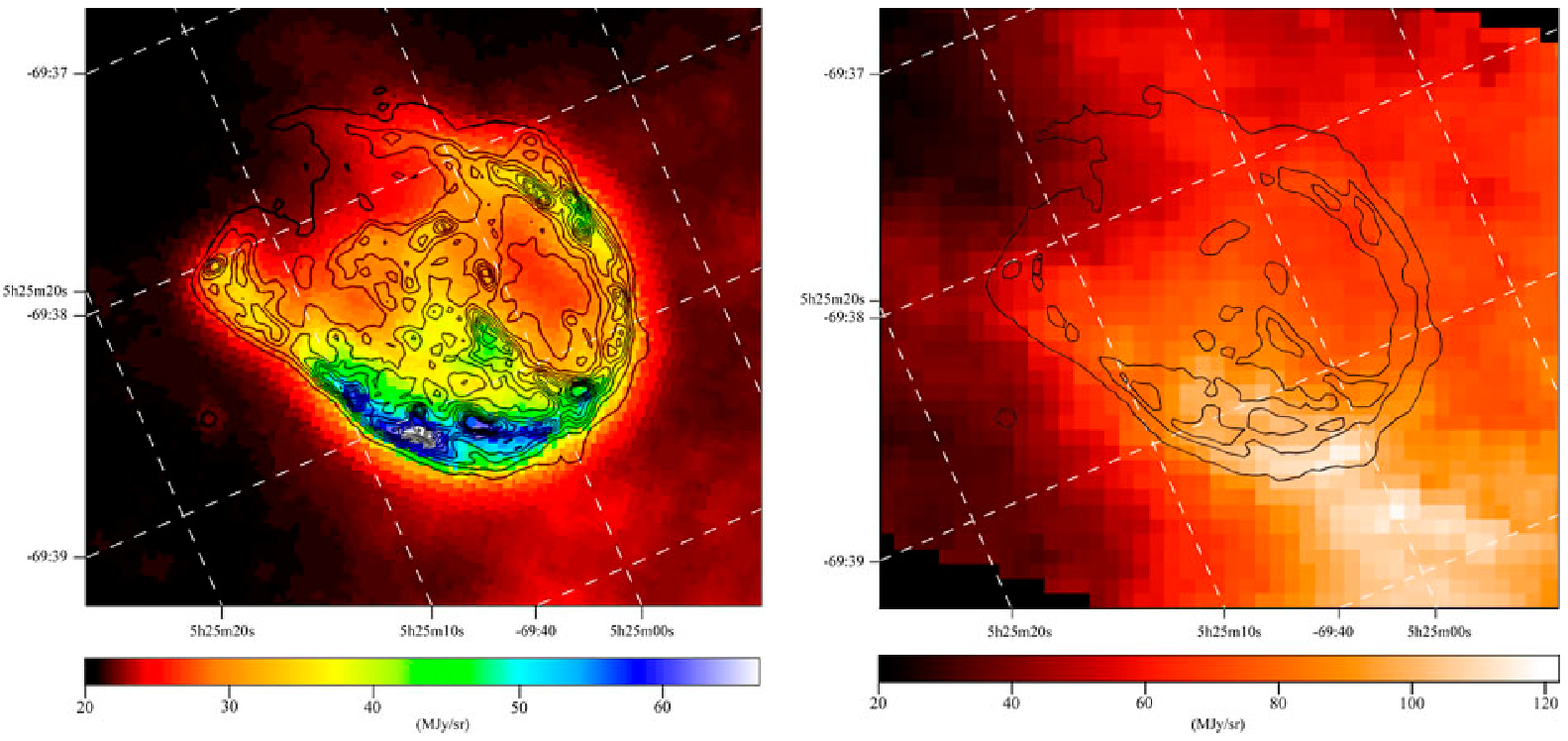} \caption{N132D: ({\it left panel}) {\it Spitzer} MIPS\,24\mic\
 image superposed with archival {\it Chandra} ACIS 0.3--10\,keV contours; note the star at $\rm 05^h25^m$16\fs7,
$-$69\arcdeg38\arcmin39\farcs9 \citep[classified as a semi-regular pulsating star by][but being a source of hard
X-ray emission]{cioni2001}, which shows the approximate width of the MIPS point-spread-function (PSF). The pixel
scale is 2.55\arcsec/pixel or 0.62\,pc/pixel; ({\it right panel}) {\it Spitzer} MIPS\,70\mic\ image superposed
with {\it Chandra} ACIS 0.3--10\,keV contours. The pixel scale is 5\arcsec/pixel or 1.21\,pc/pixel.}
 \label{fig:n132dmipsxray}
 \label{fig:n132dmips70xray}
\end{figure*}

N132D shows strong emission in the MIPS\,24\mic\ band with a background subtracted peak surface brightness of
50\mj\ in the southeastern rim. The total, background subtracted MIPS\,24\mic\ flux integrated over the whole
remnant is $3.0\pm0.2$\,Jy. The surface brightness distribution follows the X-ray contours very well
(cf.~Fig.~\ref{fig:n132dmipsxray}, left panel), although note that the MIPS\,24\mic\ emission visibly extends
beyond the edge of the remnant as defined by the X-ray contours, particularly along the western, southern, and
eastern edges. This is a clear sign of the radiative precursor also seen in the optical \citep{morse1996}, and is
most likely continuum emission from heated dust and/or line emission from ionized oxygen caused by hard photons
from the remnant; it cannot solely be explained by the wider MIPS point-spread-function (cf.~point source in
Fig.~\ref{fig:n132dmipsxray}, left panel). The remnant is also marginally detected in the MIPS\,70\mic\ band with
about 60\mj\ background subtracted peak surface brightness and a total integrated flux of $9.6\pm1.7$\,Jy, but the
background confusion is much stronger (Fig.~\ref{fig:n132dmips70xray}, right panel). There is no conclusive
detection in the MIPS\,160\mic\ band, with an upper limit of 5\mj\ peak surface brightness.
\notetoeditor{Figure f2 and f3 should appear adjacent(f2 below f3) and together on one page, preferably top
position}
\begin{figure*}[!tbp]
 \plotone{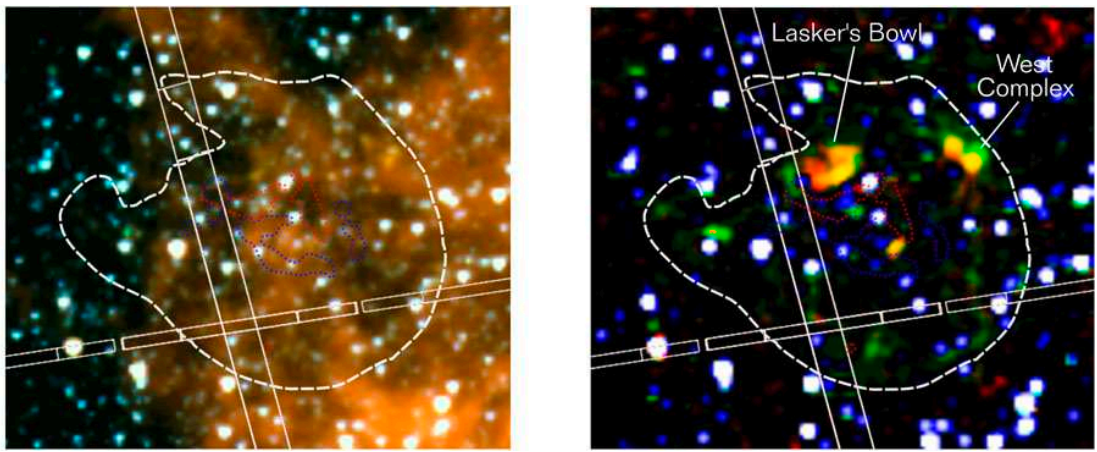} \caption{N132D IRAC emission knots: ({\it left panel}) {\it Spitzer}
 IRAC 3.2--6.4\mic\ smoothed color composite (3.6\mic=blue, 4.5\mic=green, 5.7\mic=orange). The dashed line
represents the outermost {\it ROSAT} X-ray contour from Fig.~\ref{fig:n132d}, and the polygons represent the IRS
slit coverage. The dotted contours outline the blue and redshifted ejecta of N132D as seen in the optical
\citep{morse1995}; ({\it right panel}) {\it Spitzer} IRAC smoothed color difference image: 3.6\mic=blue/white
(faint/bright stars), 4.5\mic$-$(3.6\mic$\times$0.65)=green, 5.7\mic$-$(8.0\mic$\times$0.33)$-$
(3.6\mic$\times$0.5)=red. The contrast levels have been chosen to suppress background noise fluctuations such that
only the strongest emission peaks appear against a black background.  See Sect.~\ref{sec;IRACdiscussion} for a
detailed discussion.}
 \label{fig:n132dknots}
\end{figure*}
The four IRAC bands do not show clear evidence of emission spatially correlated with the remnant as seen in MIPS
nor with the ejecta seen in the optical (cf.~Fig.~\ref{fig:n132dknots}, left panel). This is partly due to
confusion with the foreground and background ISM material. The IRAC emission within the remnant perimeter is
clumpy and shows voids that may have been created by the passing shockwave of N132D. The IRAC color ratios
measured at different positions within the remnant are consistent with PAH emission
\citep[cf.][Fig.~2]{reach2006}. We also detected several small emission knots that show IRAC colors different from
the surroundings. In order to remove background stars and confusing PAH emission, we subtracted a scaled IRAC
3.6\mic\ image from the 4.5\mic\ image, and a scaled IRAC 3.6 and 8.0\mic\ image from the 5.6\mic\ image
(cf.~Fig.~\ref{fig:n132dknots}, right panel). We determined the optimal scaling factors empirically to produce a
generally flat background. The emission knots have been observed previously in the optical with {\it HST} WFPC2
\citep{morse1995,morse1996}, and were identified as clumps of shocked, interstellar/circumstellar material
(cf.~Sect.~\ref{sec;IRACdiscussion} for a more detailed discussion). Generally, the strongest emission in all four
IRAC bands is associated with the molecular cloud/\hii-region located south of the remnant (cf.~CO contours in
Fig.~\ref{fig:n132d}).

\subsection{IRS spectroscopy}
\label{sec;IRS}
\begin{figure}[!htbp]
\epsscale{1.15}
\centering
 \plotone{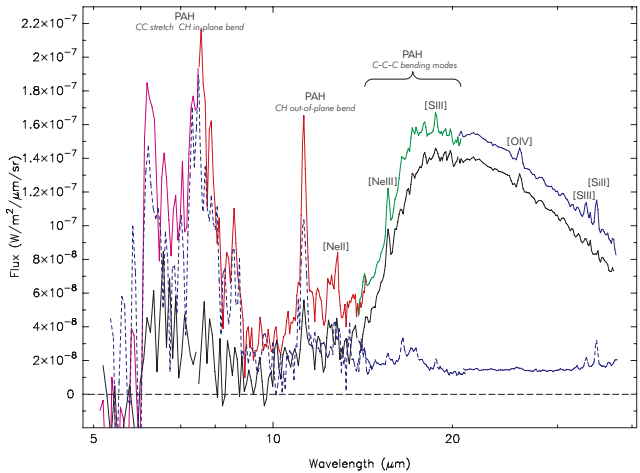} \caption{N132D southeastern rim and local background {\it Spitzer} IRS spectrum
 (online color version: magenta=SL2, red=SL1, green=LL2, blue=LL1; blue dashed=background): The upper spectrum is {\it
 without}, and the lower spectrum is {\it with} the local background subtracted (solid black lines). The
 dashed line is the local background spectrum extracted from the blue boxes in Fig.~\ref{fig:n132d}. The SL parts
 of the spectra have been slightly smoothed for illustration purposes.
 }
 \label{fig:n132dspec+backgr}
\end{figure}

\begin{figure}[!htbp]
 \epsscale{1.15}
\centering
 \plotone{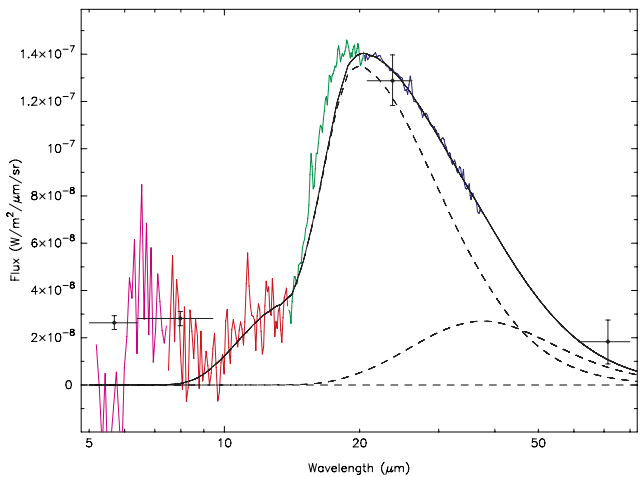} \caption{N132D southeastern rim {\it Spitzer} IRS spectrum (online color version:
 magenta=SL2, red=SL1, green=LL2, blue=LL1): diamond data points with error bars and bandwidths are background subtracted IRAC
 5.7, 8\mic\ and MIPS 24, 70\mic\ fluxes measured at the position of the extracted IRS LL spectrum. The SL parts
 have been slightly smoothed (see Fig.~\ref{fig:n132dspec-bbfit} for the original, unsmoothed version). The solid and dashed lines
 represent a two-component modified-blackbody fit to the spectrum; the fitted temperatures are 58 and 110\,K (see text for
 details).}
 \label{fig:n132dspec+bbfit}
\end{figure}

Figure~\ref{fig:n132dspec+backgr} compares the IRS spectrum extracted from the southeastern rim of N132D with the
local background spectrum.
Figure~\ref{fig:n132dspec+bbfit} shows the background subtracted IRS spectrum from the southeastern rim of N132D
together with a two-component modified-blackbody fit.
Given the difficult background removal and the slightly offset extractions of SL and LL spectra
(cf.~Sect.~\ref{sec;background}), it is worthwhile to note that the spectra from all modules seamlessly connect,
and the fluxes are consistent with background subtracted flux estimates based on the calibrated IRAC and MIPS
data.

The modified-blackbody fit to the continuum uses an emissivity for 0.1\mic\ silicate/graphite particles with a
silicate-to-graphite ratio of 0.2, following the ISM silicon and carbon elemental abundances derived for N132D by
\citet{blair2000} and silicon abundances derived from X-ray data by \citet{hughe1998} and \citet{favat1998}. The
shape of this simple emissivity function is very similar to the emissivity from the general LMC dust model of
\citet{weing2001} in the 5--35\mic\ range. Note the shoulder near 10\mic\ in the modified-blackbody fit due to the
silicate part of the emissivity, which reproduces the continuum flux of our spectrum and therefore provides direct
evidence that the heated dust in N132D contains at least some fraction of silicates. Modelling of the UV to
near-IR extinction curve for LMC sightlines \citep{pei1992,macci1994,clayt2003,cartl2005} generally seems to
indicate a larger silicate content in dust grains, i.e.~favouring larger silicate to graphite abundance ratios.
For comparison, we also used the emissivity from the LMC dust model by \citet{weing2001} and alternative
silicate-to-graphite abundance ratios of 1 and 4 for our 0.1\mic\ silicate/graphite dust emissivity. The original
emissivity data for silicate and graphite particles are taken from \citet{draine1984} and
\citet{laor1993}\footnote{Emissivity data files are available from the homepage of Bruce T. Draine, {\tt
http://www.astro.princeton.edu/$\sim$draine}}. We deliberately excluded the PAH component from the emissivity in
order to analyze these features later in the continuum subtracted spectrum (Fig.~\ref{fig:n132dspec-bbfit}). The
SMART package uses a modified-blackbody fitting function of the form
\begin{equation}\label{eqn;bbfitfunction}
{\rm{Flux~density}} = \left( {1 - e^{ - \alpha  Q_\lambda  } } \right)\sum\limits_i {\Omega _i }  B_{\lambda}
(T_{i}),
\end{equation}
where $\alpha  Q_\lambda$ is the normalized emissivity $Q$ scaled by the factor $\alpha$, and $\Omega
B_{\lambda}(T)$ is the Planck function at temperature $T$ scaled by the factor $\Omega$, which accounts for the
size of the emissive region. With the fitted free parameters $\alpha$, $\Omega$, and $T$, all emissivity functions
mentioned above produce nearly equally good fits to the spectral data, and the fitted values for $T$ and $\Omega$
are similar for all models. For a two-component fit with a silicate-to-graphite ratio of 0.2, we obtain
temperatures of $T_{1}=58\pm15$\,K and $T_{2}=110\pm6$\,K, and a ratio $\Omega_{1}/\Omega_{2}\approx15$.

\subsubsection{PAH features from the supernova blast wave}
\label{sec:discussion_PAHlines}
\begin{figure}[!tp]
\epsscale{1.15}
\centering
 \plotone{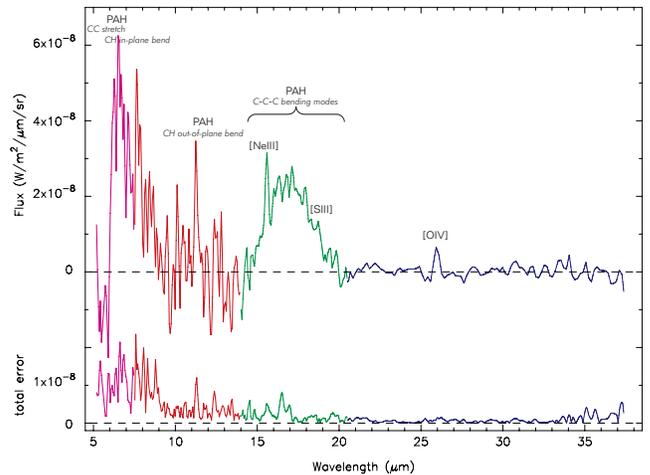} \caption{N132D southeastern rim spectrum (color online version: magenta=SL2, red=SL1,
 green=LL2, blue=LL1): the local background and the modified-blackbody continuum from Fig.~\ref{fig:n132dspec+bbfit} have
 been subtracted. The total error includes the reduction pipeline delivered errors and an estimate of
 the variation in the local background (see text for details). }
 \label{fig:n132dspec-bbfit}
\end{figure}

Figure~\ref{fig:n132dspec-bbfit} shows the IRS spectrum of the southern rim. The foreground/background and the
fitted dust continuum have been subtracted (cf.~Sect.~\ref{sec;background} and \ref{sec;IRS}). The quoted total
error consists of the pipeline-delivered statistical ramp uncertainties, combined with an estimate of the error
due to the local background variation (cf.~Sect.~\ref{sec;background}).
We detect several emission lines from atomic ions, tentative weak PAH features near 6.2 and 7.7--8.6\mic, a weak
PAH 11.3\mic\ feature, and a prominent, broad hump at 15--20\mic. The ionic emission lines are from Ne$^{2+}$,
O$^{3+}$, and S$^{2+}$ ions. We can not distinguish between \oivf\,25.91\mic\ and \feiif\,25.99\mic\ in our
low-resolution spectra, but argue that most of the emission is from O$^{3+}$ (cf.~Sect.~\ref{sec;atomiclines}).
The PAH features at 6.2, between 7.7 and 8.6\mic, and at 11.3\mic\ are the well-studied PAH C-H in-plane and
out-of-plane bending modes \citep[e.g.][]{hony2001,peete2002,vandi2004}. The presence of discrete 6.2 and
7.7--8.6\mic\ bands is uncertain due to the low quality of our spectra in that region (cf.~total error in
Fig.~\ref{fig:n132dspec-bbfit}). There is probably an underlying emission plateau at 6--9\mic, which has been
observed in other sources and is probably associated with broad continua due to large PAHs, clusters of small
PAHs, or very small grains \citep[e.g.][]{allam1989,peete2002,rapac2005}.

\begin{figure}[!htbp]
\epsscale{1.15}
\centering
 \plotone{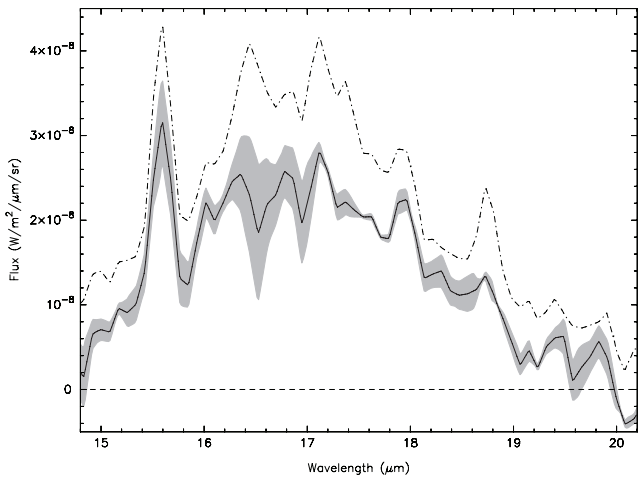} \caption{N132D southeastern rim spectrum (LL2): detailed view of 15--20\mic\ region from
 Fig.~\ref{fig:n132dspec-bbfit}. For comparison, we show the spectrum without the local background subtraction
 (dot-dashed line), and the total error from Fig.~\ref{fig:n132dspec-bbfit} as a gray-shaded error band.
 }
 \label{fig:n132dspec_gaussfit}
\end{figure}

The broad hump at 15--20\mic\ is presumably due to PAH \mbox{C-C-C} in- and out-of-plane bending modes
\citep{allam1989,vanke2000,peete2004}. Figure~\ref{fig:n132dspec_gaussfit} shows the prominent emission hump
between 15 and 20\mic\ in detail. The spectrum without background subtraction clearly shows the 16.4\mic\ PAH
feature \citep[cf.][]{mouto2000,vanke2000}, and additional features near 17.1 and 17.9\mic. These distinct bands
with widths $\sim$\,0.5\mic\ may arise from smaller PAHs in the gas phase. However, the individual bands are
hardly distinguishable after background subtraction, indicating that these features are primarily due to the
background (cf.~background spectrum in Fig.~\ref{fig:n132dspec+backgr}).

Table~\ref{tab;gaussfit} lists the fit parameters for all detected ionic lines and PAH features in the rim, the
local background, and the molecular cloud/\hii-region near N132D. For the local background spectrum, we assumed a
flat continuum of $1.2\times10^{-8}\,\rm W\,m^{-2}\mic^{-1}sr^{-1}$.
\begin{table}[!tbp]
\centering \caption{N132D fit data with errors given in parentheses: southeastern rim
(Fig.~\ref{fig:n132dspec-bbfit}), local background (Fig.~\ref{fig:n132dspec+backgr}), and the molecular cloud/{\rm
H}\,{\sc ii}-region (Fig.~\ref{fig:n132dspec_rim+knot+cloud+ngc} and \ref{fig:n132dspec_sources}). }
\begin{tabular}{lcc@{\hspace{0.3cm}}c@{\hspace{0.1cm}}c@{\hspace{0.3cm}}cc}  
  \hline\hline

       ID      &   \multicolumn{2}{c}{$\lambda$}        &  \multicolumn{2}{c}{\hspace{-0.25cm}Flux$\times 10^{-8}$}          &    \multicolumn{2}{c}{FWHM}       \\
                   &    \multicolumn{2}{c}{[$\mu$m]}   &  \multicolumn{2}{c}{\hspace{-0.25cm}[$\rm W\,m^{-2}\,sr^{-1}$]}    &   \multicolumn{2}{c}{[$\mu$m]}    \\
  \hline\hline

  \multicolumn{7}{c}{\it N132D southeastern rim}     \\
PAH\,6.2  &         &      &   \multicolumn{2}{l}{\raisebox{0.5mm}{\ \ \ \tiny ${\lesssim}$}1.8} &  &  \\

  PAH\,11.3  &      11.28 &       (0.01) &        1.06 &        (0.12) &       0.20 &       (0.05) \\

   [Ne\,{\sc iii}] & 15.60 &       (0.01) &        0.81 &        (0.01) &       0.27 &       (0.03) \\

      PAH\,15--20  &       &         &        7.47 &        (0.67) &       &        \\

[S\,{\sc iii}] 2--1&    18.74 &       (0.01) &        0.49 &        (0.04) &       0.40 &       (0.03) \\

     [O\,{\sc iv}] &    25.96 &       (0.01) &        0.30 &        (0.02) &       0.36 &       (0.02) \\

       \multicolumn{7}{c}{\it N132D local background}     \\
PAH\,6.2  &         &       &   \multicolumn{2}{l}{\raisebox{0.5mm}{\ \ \ \tiny ${\lesssim}$}5.8} &  &  \\

  PAH\,11.3  &      11.26 &       (0.01) &        4.28 &        (0.15) &       0.28 &       (0.02) \\

   [Ne\,{\sc iii}] & 15.58 &       (0.01) &        0.30 &        (0.02) &       0.23 &       (0.01) \\

      PAH\,15--20  &       &         &        3.20 &        (0.68) &       &        \\

[S\,{\sc iii}] 2--1&    18.74 &       (0.01) &        0.24 &        (0.02) &       0.30 &       (0.03) \\

[S\,{\sc iii}] 1--0&    33.52 &       (0.01) &        0.34 &        (0.01) &       0.46 &       (0.02) \\

[Si\,{\sc ii}]&    34.86 &       (0.01) &        0.77 &        (0.01) &       0.43 &       (0.01) \\

   \multicolumn{7}{c}{\it Molecular cloud/{\it H\,{\it\scriptsize II}}-region near N132D (LL only)}    \\

[Ne\,{\sc iii}]? &15.63 & (0.02) & 0.16 & (0.04) & 0.24 & (0.06) \\
PAH? &    15.92 &          (0.01) &  0.25     &       (0.05) &   0.23   & (0.05) \\
PAH\,16.4 &  16.46 &       (0.01) &   1.67     &       (0.08) &    0.29  & (0.01) \\
PAH?    &  16.75 &        (0.01) &   0.83     &       (0.13) &    0.35  & (0.04)\\
\htwo\ S(1)& 17.06       & (0.01) &   1.02    &       (0.01) &  0.27    & (0.01)\\
PAH\,17.4  & 17.38 &       (0.01) &    1.57    &       (0.11) &    0.41    & (0.02)\\
PAH\,17.8  &   17.76 &     (0.01) &   0.48     &       (0.07) &   0.32    &  (0.04)\\
PAH? &    18.27 &          (0.03) &  0.14     &       (0.04) &   0.30   & (0.09) \\

[S\,{\sc iii}] 2--1 & 18.77 & (0.01) &  0.68     &       (0.01) &  0.26    & (0.01) \\

\htwo\ S(0)&   28.25 &       (0.01) &  0.26     &       (0.01) &  0.38    & (0.01) \\

[S\,{\sc iii}] 1--0 &33.52 &(0.01) &  0.71      &       (0.01) &  0.41    &  (0.01)\\

[Si\,{\sc ii}] &   34.86 &  (0.01) &  0.97      &       (0.01) &  0.38    &  (0.01)\\
  \hline\\
\end{tabular}
\label{tab;gaussfit}
\end{table}
The detected PAH features in the southeastern rim of N132D are the first evidence of PAH molecules surviving a
supernova blast wave to our knowledge. PAH features also appear in lines-of-sight outside the remnant perimeter,
and it is necessary to make certain that the PAH emission originates from the blast wave, and not from foreground,
background, or ambient ISM material. There are three points of evidence that the PAH emission is indeed
originating from material associated with the blast wave. Firstly, we removed the foreground and background
emission by subtracting the flux from lines-of-sight just outside the blast wave perimeter
(Sect.~\ref{sec;background}). Secondly, we analyzed the spatial variation of emission line fluxes along the LL
spectrograph slit (cf.~Sect.~\ref{sec:discussion_spatialvariation}, Figure~\ref{fig:n132dspec_spatialvariation}),
which shows that the PAH features between 15 and 20\mic\ have a well-defined local flux maximum at the position of
the rim. This is expected for material that has been swept up by the blast wave and is not yet completely
destroyed by the shock. If the majority of the flux would be from unrelated foreground or background emission,
there would be no local maximum at the position of the rim. Some fraction of the detected PAHs may reside in the
radiative precursor (cf.~Sect.~\ref{sec;IRAC}), where they would be excited/processed by energetic photons
emanating from the supernova remnant. Finally, the 15--20\mic\ PAH emission changes notably when comparing
features from the supernova remnant to the adjacent molecular cloud/\hii-region and the local background
(cf.~Fig.~\ref{fig:n132dspec_rim+knot+cloud+ngc} and \ref{fig:n132dspec_sources}), which we interpret as PAH/grain
processing in N132D (cf.~Sect.~\ref{sec;IRSgrainprocessing_discussion} and \ref{sec:previousobservation}).

\subsection{Spatial variation of the emission features}
\label{sec:discussion_spatialvariation}
\label{sec;atomiclines}
\label{sec:PAHlines}
\begin{figure}[!tbp]
\epsscale{1.15}
\centering
 \plotone{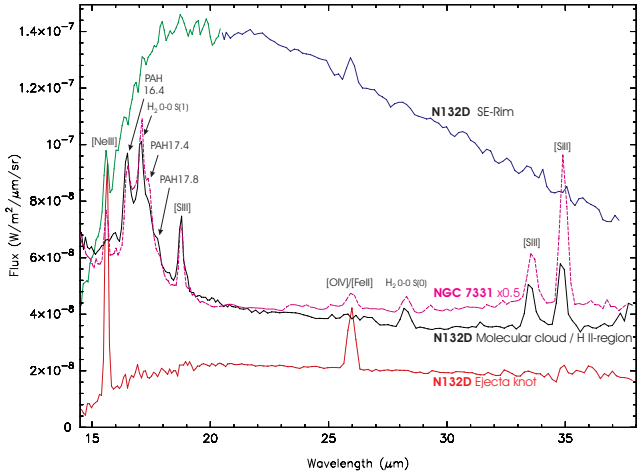} \caption{Comparison of different {\it Spitzer }IRS spectra for N132D, its nearby
molecular cloud/H\,{\sc ii}-region, and an archival spectrum of the galaxy NGC\,7331 (dashed line). We extracted
and processed the average spectrum of the nucleus and inner ring of NGC\,7331 \citep[][]{smith2004} in the same
way as our spectra, and applied a scaling factor of 0.5 for illustration purposes (see
Sect.~\ref{sec:discussion_spatialvariation} and \ref{sec:previousobservation} for further discussion). }
 \label{fig:n132dspec_rim+knot+cloud}
 \label{fig:n132dspec_rim+knot+cloud+ngc}
\end{figure}

In Figure~\ref{fig:n132dspec_rim+knot+cloud}, we show the spectrum of the southeastern rim of N132D in comparison
to the ejecta knot detected in the LL spectra, the molecular cloud/\hii-region south of N132D, and the average
spectrum of the galaxy NGC\,7331. The characteristic features of the rim spectrum have been described earlier in
this paper. The spectrum of the ejecta knot lacks the steeply rising dust continuum and the 15--20\mic\ PAH
features, but shows strong ionic emission lines from Ne$^{2+}$ and O$^{3+}$, which is consistent with the strong
neon and oxygen lines seen in the optical for the same knot \citep[N132D-P3,][]{blair2000}. The line fluxes and
FWHM are $\rm 1.6\times10^{-8}\wms$ (0.15\mic) for the neon and $\rm 8.4\times10^{-9}\wms$ (0.34\mic) for the
oxygen line. The spectrum of the molecular cloud/\hii-region shows strong PAH bands as well as ionic and \htwo\
lines. The association of the CO emission south of N132D with an \hii-region seen in H$\alpha$ was noticed by
\citet{morse1995}. \citet{banas1997} estimated the mass of the CO cloud to be about $3\times10^5\msun$, and argued
that its spatial proximity and matching LSR velocity provide strong evidence that this giant cloud is associated
with N132D. This provides an opportunity to relate our observations of dust and PAHs in the supernova remnant
N132D to the nearby environment its progenitor star probably originated from. Such a comparison will also be
useful to interpret spectroscopic observations of star-forming galaxies \citep[e.g.~NGC\,7331,][see
Sect.~\ref{sec:previousobservation}]{smith2004}.

\begin{figure}[!htbp]
\epsscale{1.15}
\centering
 \plotone{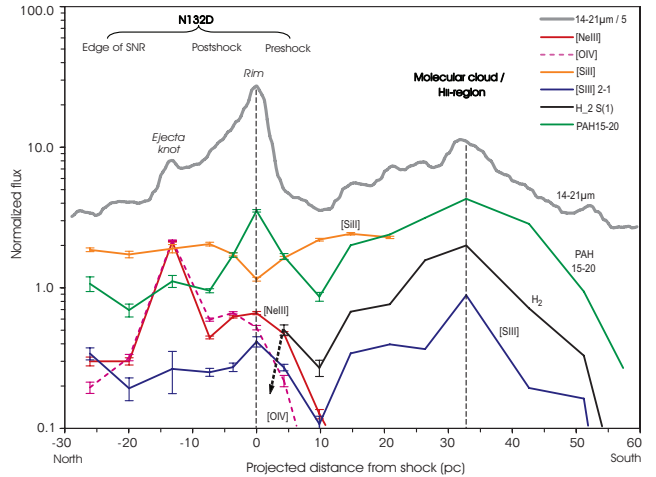} \caption{Spatial variation of line vs.~PAH 15--20\mic\ fluxes, and total integrated flux from continuum and lines in
 the 14--21\mic\ range as a function of projected distance from the supernova shock. [Si\,{\sc ii}] and [O\,{\sc
 iv}] appear only in LL1 (19.5--38\mic), limiting the spatial coverage to about 20\,pc. The \htwo\ flux across the
 supernova remnant is uncertain due to strong blending with PAH features but most likely very small (see text for
 further discussion).
  }
 \label{fig:n132dspec_spatialvariation}
\end{figure}

We present a systematic assessment of the PAH 15--20\mic, ionic, and molecular line fluxes from positions across
N132D and the nearby molecular cloud in Figure~\ref{fig:n132dspec_spatialvariation}. We extracted spectra along
the LL slit with $\approx25\arcsec$ separation starting at the northern edge of N132D and ending at the southern
edge of the molecular cloud about 3.3\arcmin\ south of N132D, resulting in a total number of 15 spectra. We
converted the spatial distance along the slit into a parsec scale assuming a LMC distance of 50\,kpc, and adopted
the location of the southeastern rim of N132D as the zero position. Each node in
Figure~\ref{fig:n132dspec_spatialvariation} represents a line flux derived from continuum and line fitting, with
errors given for the positions covering the supernova remnant (cf.~Sect.~\ref{sec:analysis_emissionlines}).

The ionic and molecular lines detected and examined in our spectra are \neiiif\,15.6,  \siiif\,18.7 and 33.5,
\oivf\,25.9, \siliif\,34.8, and \htwo\,0--0\,S(0)\,28.2 and S(1)\,17.0\mic. The lines from atomic ions,
particularly the strong \neiiif\  and \siiif\ transitions, are mostly unblended and straightforward to fit, which
makes them excellent references in terms of flux and central wavelengths.

\neiiif\ and \oivf\ lines only appear across the supernova remnant and in the precursor, and are particulary
strong at the position of the ejecta knot. Generally, \neiiif\ and \oivf\ trace the hot ionized gas of N132D. Note
that we can not distinguish between \oivf\,25.91\mic\ and \feiif\,25.99\mic\ in our low-resolution spectra, but we
argue that most of the flux is likely to be from \oivf\ because the flux follows \neiiif\ closely
(cf.~Fig.~\ref{fig:n132dspec_spatialvariation}); under the conditions where we observe Ne$^{2+}$, iron would be
mostly present as Fe$^{3+}$ and Fe$^{4+}$. There is a weak line near 17.9\mic, which may be \feiif\,17.9, but
possible confusion with the PAH 15--20\mic\ features renders this line inconclusive. We do not see
\feiif\,24.5/35.3\mic, \feiif/\feiiif\,22.9, and \fevif\,19.6\mic, which are probably too weak to be detected in
our spectra. However, X-ray emission from Fe$^{16+}$ and more highly ionized states was detected in the southern
rim \citep{behar2001}.

\siiif\ is a common line seen in PDRs and \hii-regions. Here, it not only appears across the remnant, but also
across the molecular cloud tracing the \hii-region. The \siiif\,1-0 transition at 33.5\mic\ follows the same
general trend, with the \mbox{2-1/1-0} ratio increasing from 0.08 to about 0.2 from the northern edge of the
molecular cloud into the remnant.

\htwo\ traces the  molecular cloud seen in CO (Fig.~\ref{fig:n132d}). We checked for \htwo\,0-0\,S(0)\,28.2,
S(2)\,12.3, and S(3)\,9.7\mic, and did not detect these lines across the remnant; weak S(1)\,17.0\mic\ emission is
not detectable because of strong blending with PAH emission. We conclude that \htwo\ is mostly absent across the
remnant, where it was probably destroyed by the fast shock and the strong radiation field, but apparently present
in the precursor. The \htwo\,S(1)/S(0) ratio increases from about 1.4 at the northern edge of the molecular
cloud/\hii-region to \raisebox{0.5mm}{\scriptsize ${\gtrsim}$}\,11 at the shock precursor of the rim, which
implies an increase in the UV radiation field as expected for the presence of a radiative precursor
(cf.~Sect.~\ref{sec;IRAC}).

\siliif\ shows a nearly constant flux, except a noticeable minimum at the position of the rim. This is most likely
an effect of the increased radiation field and the elevated temperature. Silicon will be much more highly ionized
under those conditions, which is confirmed by the detection of X-ray emission from Si$^{12+}$ in the southern rim
by \citet{behar2001}. Unfortunately, this conceals a possibly increased amount of silicon in the gas phase due to
the destruction of dust grains.

There are three well-defined PAH components at 16.5, 17.4, and 17.8\mic\ in the region of the molecular cloud
(Fig.~\ref{fig:n132dspec_rim+knot+cloud+ngc} and \ref{fig:n132dspec_sources}, and Table~\ref{tab;gaussfit}), which
have all been previously observed in other sources (see Sect.~\ref{sec:previousobservation}). The fluxes of these
features follow the \htwo\ flux very well as expected if the PAHs are well-mixed with the gas
(Fig.~\ref{fig:n132dspec_spatialvariation}). The feature at 16.8\mic\ is blended with the stronger PAH feature at
16.5\mic\ and the strong \htwo\,S(1) line at 17.1\mic, and therefore more uncertain. The situation changes for the
supernova remnant, where the discrete 15--20\mic\ PAH features all but disappear, and a nearly featureless
15--20\mic\ emission hump remains (Fig.~\ref{fig:n132dspec_gaussfit} and Table~\ref{tab;gaussfit}).


\section{Discussion}
\label{sec;discussion}
\subsection{IRAC emission knots}
\label{sec;IRACdiscussion}
We detect shocked interstellar clouds with IRAC due to their distinct emission in the IRAC channels
(cf.~Fig.~\ref{fig:n132dknots}). The IRAC emission knots seen in the northern central part and the northwestern
rim of the remnant coincide with previously detected, bright optical \oiiif\ and \siif\ emission complexes named
``Lasker's Bowl'' and ``West Complex''. They were identified as shocked interstellar clouds/circumstellar material
based on optical line diagnostics, elemental abundances, and their radial velocities similar to the local ISM
surrounding N132D \citep[][]{morse1995,morse1996,blair2000}. The small knot visible close to the center of the
remnant is probably a smaller analogue of these two complexes. We do not detect the red- and blueshifted,
oxygen-rich ejecta of N132D with IRAC (cf.~the dotted contours in Fig.~\ref{fig:n132dknots}).

The measured IRAC band ratios for the shocked clouds are roughly consistent with thermal emission from hot
silicate/graphite dust grains. In particular, the West Complex shows an IRAC 4.5/8\mic\ ratio of 0.18 and a
3.6/5.7\mic\ ratio of 0.38. For this region, \citet{blair2000} presented {\it Hubble} FOS spectra that show a
broad range of ionization stages as expected from shock heating, e.g.~O to O$^{3+}$ and N to N$^{4+}$, and
elemental abundances consistent with LMC interstellar material. They derive a high electron temperature
$T\approx43,000$\,K, a mean electron density of several thousand per \ccm, and a preshock cloud density larger
than 100\ccm. Their shock models indicate that a shock speed of at least $\sim$\,200\km\ is needed to produce the
observed high-ionization species.

Comparing this to shocked dust IR emission models calculated by \citet{dwek1996}, we find that the measured IRAC
color ratios are roughly reproduced by models with a preshock density of 100\ccm\ and a shock speed
\raisebox{0.5mm}{\scriptsize $\lesssim$}\,400\km, and models with a preshock density of 10\ccm\ and a range of
shock speeds between 400 and 1000\km. This suggests that part of the IRAC emission in the West Complex may be due
to a population of shock-heated, small dust grains. However, it is likely that the measured IRAC colors represent
a mix of dust continuum and ionic emission. The green channel 2 excess emission probably includes
\hi\,Br$\alpha$\,4.05\mic, since the knots appear in the H$\alpha$ image from \citet{morse1995} and the optical
spectrum of the West Complex shows strong hydrogen recombination lines \citep{blair2000}. The red channel 3 excess
emission could also be due to \feiif\,5.34\mic. If \htwo\ is present in those clouds, the IRAC 4.5\mic\ band
covers the more highly excited lines, i.e.~the \htwo\,(0,0) S(8)-S(11) transitions, whereas the 5.7 and 8\mic\
bands cover \htwo\,(0,0) S(4)-S(7).

A different type of IRAC emission is mainly characterized by an excess in channel 2, i.e.~it appears as pure green
emission in Fig.~\ref{fig:n132dknots}, right panel. The faint green emission aligned along the southwestern rim is
spatially coinciding with small clumps of gas seen in the {\it HST} WFPC2 images \citep{morse1996}. Judging solely
from the IRAC band fluxes, it is difficult to determine the nature of the IR emission in these clumps.
Unfortunately, none of them are covered by our IRS spectra. Many other isolated patches showing channel 2 excess
emission do not have a counterpart in the narrow band {\it HST} WFPC2 \oiiif, \siif, nor the H$\alpha$ image from
\citet{morse1995}. However, the brightest green patches, one near the center of the ejecta, two blended patches
near the northeastern edge of the remnant, and one outside the remnant perimeter near the northeastern corner of
the image, have optical counterparts visible in archival, broad band {\it HST} ACS data (proposal 12001, filters
F475W, F550M, F850LP; see the related Hubble Heritage image, STScI-2005-30 news release). All counterparts are
structured, extended sources with color variations. The object outside the remnant perimeter is most likely a
background galaxy, whereas the nature of the objects within the perimeter of N132D remains unclear.

\subsection{MIPS continuum emission from shock-heated grains}
\label{sec;MIPSdiscussion}
We interpret the steeply rising continuum in the spectrum of the southeastern rim of N132D as thermal emission
from dust grains that have been swept up and shock-heated by the supernova blast wave. Most of the MIPS\,24\mic\
flux is then due to dust continuum flux, since the \oivf\,25.9\mic\ line flux is much weaker than the continuum in
the wide MIPS\,24\mic\ bandpass (cf.~Fig.~\ref{fig:n132dspec+bbfit}). The association of the MIPS\,24\mic\ flux
with the blast wave along the remnant perimeter and not with the actual ejecta (cf.~Fig.~\ref{fig:n132dmipsxray}
and \ref{fig:n132dknots}) confirms that the dust is part of the swept-up, ambient interstellar/circumstellar
material. This is further supported by the appearance of the brightest MIPS emission along the southeastern rim,
where the interstellar gas density is expected to increase due to the presence of a molecular cloud (cf.~CO
contours in Fig.~\ref{fig:n132d}, and Sect.~\ref{sec:discussion_spatialvariation}). Generally, the MIPS\,24\mic\
flux intensity follows the X-ray emission very well. Every peak in the MIPS\,24\mic\ surface brightness is also an
X-ray peak (cf.~Fig.~\ref{fig:n132dmipsxray}), which demonstrates the connection between the X-ray emitting plasma
and the heated dust grains.

A two-component modified-blackbody fit to the continuum yields temperatures of 110 and 58\,K
(cf.~Sect.~\ref{sec;IRS}). These temperatures can be adopted as average dust temperatures if the grains are in
thermal equilibrium, i.e.~balanced heating and cooling rates, and if the mean energy per grain heating event is
much smaller than the thermal energy content of the grain at the equilibrium temperature \citep[steady
vs.~stochastic heating, cf.][ and references therein]{draine2001}. Under the conditions in the rim of N132D,
i.e.~an average temperature $T\sim8.4\times10^6$\,K and a mean preshock density \raisebox{0.5mm}{\scriptsize
$\gtrsim$}\,3\ccm\ \citep[cf.][]{hwang1993,morse1996}, gas-grain collisions probably dominate the grain heating
\citep{dwek1992}. If each gas-grain collision transfers an energy of the order of ${\rm
k}T\approx1\times10^{-16}$\,J, i.e.~similar to the energy of a 0.6\,keV X-ray photon, an individual heating event
changes the temperature of a silicate/graphite grain with radius \raisebox{0.5mm}{\scriptsize
$\gtrsim$}\,0.04\mic\ by less than 1\,K. This estimate assumes spherical, homogenous grains with a density of $\rm
2500\,kg\,m^{-3}$, and a specific heat capacity of $\rm 185\,J\,kg^{-1}K^{-1}$ for silicate/graphite grains at
$T\sim100$\,K. Small grains \raisebox{0.5mm}{\scriptsize $\lesssim$}\,0.04\mic\ in size can be heated
substantially by a single heating event, and thus do not have a well-defined equilibrium temperature
\citep[cf.][]{li2001}. The fact that we observe two different average grain temperatures may indicate two distinct
grain size populations and/or emission originating from separate physical regions with different heating
conditions (see~Sect.~\ref{sec;IRSgrainprocessing_discussion} for a more detailed discussion).

We can estimate the dust mass $M_{\rm  dust}$ in those two components using \citep{whitt2003}
\begin{equation}\label{eqn;dustmass}
M_{{\rm{dust}}}  = \frac{{4\rho F_\lambda  d^2 }}{{3B_\lambda  (T_{{\rm{dust}}} )}}\left[ {\frac{a}{{Q_\lambda }}}
\right],
\end{equation}
and adopting a density $\rm \rho=2500\,kg\,m^{-3}$ for silicate/graphite grains, a distance $d=50 \rm \,kpc$ for
N132D, and an average ratio of grain radius $a$ over emissivity $Q_\lambda$ of $1\times10^{-5}\rm m$. We evaluated
the flux density $F_\lambda$ and the Planck function $B_\lambda$ for the fitted dust temperatures $T_{\rm dust}$
at 23.7\mic. Eq.~\ref{eqn;dustmass} is an approximation assuming spherical dust grains of uniform size,
composition, and in thermal equilibrium. We derived the dust masses in the entire southern rim (outlined by the
green color in Fig.~\ref{fig:n132dmipsxray}, left panel) by scaling the result from Eq.~\ref{eqn;dustmass} by the
MIPS\,24\mic\ flux ratio of the southern rim and the IRS slit aperture, which gives
$M_{{\rm{dust,\,58\,K}}}\sim0.05\rm\,M_\odot$ and $M_{{\rm{dust,\,110\,K}}}\sim0.003\rm\,M_\odot$. If we further
scale this result to the total MIPS\,24\mic\ flux of the whole remnant, the total swept-up ISM dust mass in N132D
becomes about $\rm 0.13\,M_\odot$. These numbers have considerable uncertainty, particularly the low temperature
component, because of the assumptions made in Eq.~\ref{eqn;dustmass} and the measurement uncertainties.

We can compare the derived dust masses with estimates of the swept-up gas mass from X-ray data, which give an
approximate upper limit on the total amount of dust. \citet{favat1997} fitted the BeppoSAX X-ray spectrum of N132D
with a two-temperature non-equilibrium ionization model. Using the revised best-fit from \citet{favat1998}, the
total mass of the  X-ray emitting material is about 380--850\msun, assuming a filling factor between 0.1 and 0.5.
\citet{hughe1998} applied Sedov models to fit {\it ASCA} spectra of N132D, and derived a swept-up mass of about
650\msun. Assuming a total swept-up gas mass of 600\msun\ for N132D, the amount of swept-up dust is
$\sim$\,1.5\msun, if the fraction of dust is 0.25 percent of the gas mass, i.e.~about one fourth of the Galactic
value. This dust mass represents an approximate upper limit for the case of no dust destruction. Considering the
factor-of-ten lower dust mass estimated directly from Eq.~\ref{eqn;dustmass}, we can state that although some
portion of the dust in N132D is surviving, as much as 90 percent might have been destroyed by the blast wave shock
(cf.~Sect.~\ref{sec;IRSgrainprocessing_discussion}).

\subsection{Grain processing and destruction in N132D}
\label{sec;IRSgrainprocessing_discussion}
\label{sec:Dwek_IRmodel_spectra}
Strong supernova shock waves compress, heat, and accelerate interstellar material, creating physical conditions
that destroy interstellar dust grains. The grains are not instantly following the gas acceleration due to their
inertia, but are gradually accelerated via gas-grain collisions, which cause a gradual grain erosion through
sputtering \citep[cf.][]{tiele1994}. Furthermore, the grain population develops velocity differences, and
subsequent grain-grain collisions cause shattering and vaporization depending on the relative velocity and grain
sizes \citep{tiele1994,borko1995,jones1996}.

For N132D, \citet{morse1996} have estimated a mean shock velocity of the X-ray emitting main blast wave of
$\sim$\,800\km, a mean preshock density of $\sim$\,3\ccm, and a postshock temperature of $8.4\times10^6$\,K
\citep[cf.~also][]{hwang1993}, which is in good agreement with the Sedov model values from \citet{hughe1998}. For
such a high-velocity shock, the dominant grain destruction process is thermal sputtering
\citep{dwek1992,tiele1994,borko1995}. The lifetime $\tau$ of silicate/graphite grains against this process was
estimated by \citet{draine1979}, and further refined by \citet{tiele1994},
\begin{equation}\label{eqn;thermsput}
\tau  = \left. {\begin{array}{l} 0.79\ {\rm for\ graphite}\\ 2.25\ {\rm for\ silicate} \end{array}} \right\}\times
10^4 {\rm{yr}}\left( {\frac{{{\rm{cm}}^{{\rm{
- 3}}} }}{{n_{\rm{H}} }}} \right)\left( {\frac{a}{{0.01{\rm{\mu m}}}}} \right)
\end{equation}
at a temperature of $8.4\times10^6$\,K. Thus, if the hydrogen density is $n_{\rm H}=10\ccm$, silicate/graphite
grains with a radius $a=0.01\mic$ are completely eroded after roughly 1500\,yr. Likewise, 0.001\mic\ grains only
survive 150\,yr, whereas 0.1\mic\ grains would survive 15,000\,yr.

\citet{hughe1987}, \citet{morse1996}, and \citet{hughe1998} proposed that the supernova of N132D exploded most
likely within a low-density cavity in order to reconcile the Sedov age with the dynamical age without invoking
large explosion energies. This scenario would allow the blast wave to expand unhindered in the beginning until it
encountered the denser material at the boundary of the cavity. If we assume an initial, average expansion velocity
of $\sim\!10^4$\km, the shock reached the cavity wall within about 800 years, supposing that the edge of the
cavity lies close to the current MIPS\,24\mic\ rim in the south of N132D and adopting the common center of the
oxygen-rich ejecta as the expansion origin \citep[cf.][]{morse1995}. Since the dynamical age of N132D is
$\sim$\,2500\,yr \citep[cf.][]{morse1995}, we can infer that the shock has been interacting with the denser
material for maybe 1700\,yr. Hence, we would expect that most of the grains \raisebox{0.5mm}{\scriptsize
${\lesssim}$}\,0.01\mic\ that were {\it initially} present are destroyed by now through thermal sputtering {\it
if} they were continuously residing in the hot plasma (cf.~Eq.~\ref{eqn;thermsput}). Naturally, as the shock wave
keeps expanding into new ISM material, the grain population is continually replenished.

In addition to thermal sputtering, shattering and vaporization remove some fraction of the large-size grains upon
grain-grain collision. This is limited by the probability of a grain-grain encounter before being brought to rest
with respect to the gas, which is of the order of the dust-to-gas mass ratio in the absence of betatron
acceleration \citep{draine1993,Draine2004}. Therefore, we conclude that thermal sputtering generally dominates the
grain destruction in the fast shock of N132D, but shattering may redistribute the grain size distribution
appreciably in denser clumps \citep[cf.][]{jones1996}, where the shock speed may be much lower than the estimated
mean value of $\sim$\,800\km.

\begin{figure}[!htbp]
\epsscale{1.15}
\centering
 \plotone{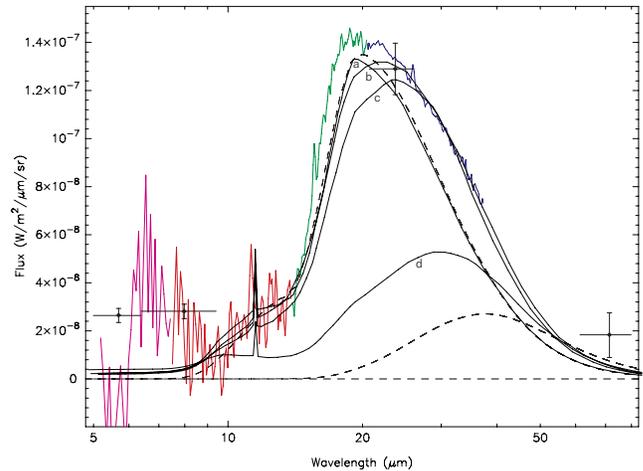} \caption{N132D southeastern rim spectrum with modified-blackbody fit components (dashed
lines, cf.~Fig.~\ref{fig:n132dspec+bbfit}) and calculated shock models \citep[][solid lines]{dwek1996}. The models
are for 1500/10 (a), 1000/10 (b), 800/10 (c), and 400\km/10\ccm\ (d) shock
 speed/preshock density, and are scaled by factors of 0.13, 0.14, 0.21, and 0.5, respectively.}
 \label{fig:n132dspec_dwekmodels}
\end{figure}

We compare our results to calculated model spectra from dust grains processed in fast, nonradiative shocks by
\citet{dwek1996}. Figure \ref{fig:n132dspec_dwekmodels} shows the spectrum from the southeastern rim of N132D
together with our modified-blackbody fit components and a series of calculated shock model spectra adapted from
\citet{dwek1996}. Their model assumes silicate/graphite grains in a power-law size distribution after
\citet{mathi1977} and grain destruction solely by sputtering without shattering/vaporization.

Overall, the scaled model spectra reproduce the observed spectrum of N132D remarkably well considering the model
parameters are not fitted to the data. The applied scaling factors probably account for the filling factor of the
emitting material, and were chosen empirically to match the flux of N132D. The main difference between the
observed spectrum and the models is a lack of continuum emission at 6--9\mic\ and around 70\mic, although the
MIPS\,70\mic\ flux measurement does not provide a very stringent constraint due to the background confusion. The
long-wavelength emission may be due to high density material, e.g.~a swept-up dense knot, with a shock velocity
lower than 400\km. The lack of 6--9\mic\ flux in the models confirms that this emission is due to PAHs and not
from a population of small, heated silicate/graphite grains. Note that the narrow feature near 11.5\mic\ in the
calculated spectra is not the 11.3\mic\ PAH C-H bending mode but an IR active mode of crystalline graphite
\citep[cf.][]{drain1984}, which would likely be smoothed out in an imperfect polycrystalline sample
\citep{li2001}.

Our high-temperature modified-blackbody fit component is very similar to the 1500\km/10\ccm\ shock model, which
justifies that our continuum fit is physically meaningful, particularly for the PAH C-C-C emission in the
15--20\mic\ range (Fig.~\ref{fig:n132dspec-bbfit}). Generally, however, a single preshock density and/or a single
shock velocity are apparently not sufficient to explain the continuum of the observed spectrum. The medium into
which the shock wave expands is clumpy \citep[cf.~{\it HST} WFPC2 images of N132D by][]{morse1996}, so we expect a
combination of different densities and shock speeds.

\subsubsection{Comparison with previous observations}
\label{sec:previousobservation}
\begin{figure}[!tb]
\epsscale{1.15}
\centering
 \plotone{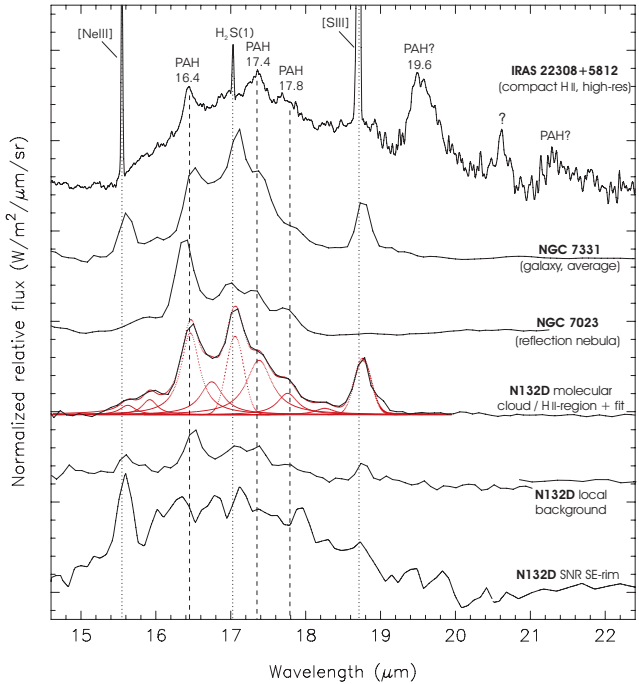} \caption{Comparison of continuum subtracted 15--22\mic\ {\it Spitzer} IRS emission
 spectra. All spectra are normalized to their 16.4\mic\ flux, and shifted along the flux axis for illustration
 purposes. We show our fit for the molecular cloud/H\,{\sc ii}-region south of N132D (online color version: red;
 cf.~Table~\ref{tab;gaussfit}). IRAS\,22308+5812 (compact H\,{\sc ii}, high-resolution mode, $R\sim600$) has a
 unique, broad feature near 19.6\mic, which also appears in {\it ISO} SWS data presented by \citet{vanke2000}, and
 weak emission longward of 20\mic. See Table\,\ref{tab:pahlineratios} for references.}
 \label{fig:n132dspec_sources}
\end{figure}

Figure~\ref{fig:n132dspec_rim+knot+cloud+ngc} shows the spectrum of the galaxy NGC\,7331 \citep[cf.][]{smith2004}
in comparison to our spectra for N132D. The spectrum of NGC\,7331 bears a striking similarity to the spectrum of
the molecular cloud/\hii-region south of N132D, and also shows \neiiif\ and \oivf/\feiif\ lines. Since it
represents an average galactic ISM mixture, the resemblance to our spectra is not surprising. \citet{smith2004}
have fitted the 15--20\mic\ PAH features in this spectrum with two Drude profile components at 16.40 ($\rm
FWHM=0.34\mic$) and 17.12\mic\ ($\rm FWHM=0.96\mic$), and measured a $16.4+17.1$ PAH flux of
$1.5\times10^{-7}$\wms. This fit reproduces the overall shape of the emission in this region together with the
\htwo\,S(1) component, but it does not take the blended features near 17.4 and 17.8\mic\ into account, which are
also observed in other sources (cf.~Fig.~\ref{fig:n132dspec_sources}). \citet{smith2004} reported the strong
17.1\mic\ feature as a possible new PAH C-C-C bending mode detection. If we adopt Lorentzian components of
narrower width, we find the same PAH components with similar fluxes and widths as seen in our spectra of the
molecular cloud/\hii-region near N132D (cf.~Table~\ref{tab;gaussfit} and Fig.~\ref{fig:n132dspec_sources}), and no
need for a broad feature near 17.1\mic. The 0.2--0.4\mic\ widths of these features are plausible for PAH molecules
in the gas phase \citep[][]{peete2004}. Our refitted, total PAH 15--20\mic\ flux in NGC\,7331 is
$1.1\times10^{-7}$\wms, which is a factor of 1.6 larger than observed in the southeastern rim of N132D, whereas
the PAH band at 11.3\mic\ is about a factor 30 stronger in NGC\,7331.

\citet[][hereafter VHP]{vanke2000} have measured 15--20\mic\ PAH plateau strengths as well as the PAH bands at 6.2
and 11.3\mic\ for a variety of Galactic sources from {\it ISO} SWS spectra. Table~\ref{tab:pahlineratios} shows a
comparison of different PAH band ratios from the {\it ISO} results by \citetalias{vanke2000} together with
reanalyzed {\it Spitzer} archival data and our results for N132D. A comparison of the absolute fluxes is hampered
by the larger {\it ISO} beam and different source distances, so we restrict ourselves to band ratios using \wms\
units instead. Note also that \citetalias{vanke2000} measured the 16.4\mic\ PAH features with a local continuum,
i.e.~just the excess portion of the feature on top of the 15--20\mic\ emission hump, whereas we generally use the
mostly emission-free flux below 15 and above 22\mic\ to determine the continuum for fitting the 15--20\mic\
features. We analyzed {\it Spitzer} archival calibration data for two sources in the \citetalias{vanke2000}
sample, IRAS\,22308+5812 and NGC\,7027, in order to test whether the results are comparable. We find that the
measured PAH ratios for these two sources match within 30 percent, which is a reasonable agreement given the
uncertainties due to the different beam sizes and continuum choices for the fitting.

\citetalias{vanke2000} interpret the observed relative strengths of the broad plateaus and the narrower PAH bands
in terms of variations in the size distribution, i.e.~the larger PAHs with a few thousand carbon atoms that give
rise to the 15--20\mic\ emission probably originated from dense cloud cores that formed the massive stars in the
\hii-regions, whereas smaller PAHs with less than a few hundred carbon atoms give rise to strong 6.2--11.3\mic\
bands. They also tentatively attribute the discrete 16.4\mic\ band to \mbox{C-C-C} bending modes of pendant
benzene ring substructures in smaller PAH molecules \citep[cf.~also][who propose PAH molecules containing
pentagonal rings as the carrier of the 16.4\mic\ band]{mouto2000}.

\begin{table}[!htbp]
\centering \caption{PAH emission band ratios in different types of sources.}

\begin{tabular}{lccccc}
\hline\hline\\[-0.3cm]
Source &                   Type    & $\frac{{15-20\mic}}{{11.3\mic}}$ &  $\frac{{16.4\mic}}{{11.3\mic}}$  & $\frac{{15-20\mic}}{{6.2\mic}}$& $\frac{{16.4\mic}}{{6.2\mic}}$ \\ 
\hline\hline\\[-0.3cm]
LkH$\alpha$\,234 \htwo-ridge &            PDR &          0.2 &       0.08  &        0.1 &  0.04  \\

NGC\,7331 (average) &                  Galaxy &        0.3 &       0.07  &        0.2 &   0.05 \\

NGC\,7023 &                  RN &          0.3 &       0.14  &        0.1 &    0.06 \\

NGC\,7027 (edge) &             PN &        0.4 &       0.04  &        1.2&   0.13  \\

*NGC\,7027 &                     PN &        0.5 &       0.04  &        0.7&  0.06   \\

*CD-42$\arcdeg$11721 &                    YSO &        0.7 &       0.10  &        0.6&  0.08   \\

*IRAS 03260+3111 &                  HAeBe &        1.0 &       0.06  &        0.8 &  0.05  \\

*IRAS 23133+6050 &                   cH\,{\sc ii} &        2.2 &     $-$     &    1.5         &  $-$  \\

*IRAS 22308+5812 &                   cH\,{\sc ii} &        2.3 &     $-$      &     1.8         &  $-$ \\

IRAS 22308+5812 &                   cH\,{\sc ii} &        2.5 &    0.27         &        0.8  &  0.08 \\

*IRAS 18317-0757 &                   cH\,{\sc ii} &        2.5 &   $-$        &    1.2         &  $-$ \\

*Sh\,2-106 IRS4 &            H\,{\sc ii} &        5.1 &     $-$       &      3.9         & $-$ \\

\hline\\[-0.3cm]

N132D SE-rim &                     SNR &        7.1$\pm$1.0 & $-$ & \raisebox{0.5mm}{\scriptsize ${>}$ }4 &   $-$  \\

N132D local backgr. &              &        0.7$\pm$0.2    &    0.18         &     \raisebox{0.5mm}{\tiny ${\gtrsim}$ }0.5     &  \raisebox{0.5mm}{\tiny ${\gtrsim}$ }0.13 \\


\hline\\[-0.1cm]
\end{tabular}
\tablecomments{Sources denoted by an asterisk are {\it ISO} SWS measurements taken from \citetalias{vanke2000} and
converted to \wms. The other sources are {\it Spitzer} archival data processed and fitted in the same manner than
N132D (NGC\,7331, see \citet{smith2004}; LkH$\alpha$\,234, see \citet{morri2004}; NGC\,7023, see
\citet{werne2004}; NGC\,7027 and IRAS\,22308+5812, {\it Spitzer} calibration observations). Dashes indicate that a
discrete 16.4\mic\ PAH band is not detected in that source. PDR\,=\,photodissociation region; RN\,=\,reflection
nebula; PN\,=\,planetary nebula; YSO\,=\,young stellar object; HAeBe\,=\,Herbig Ae/Be star; (c)H\,{\sc ii}
\,=\,(compact)H\,{\sc ii}-region.}
\label{tab:pahlineratios}
\end{table}

Table~\ref{tab:pahlineratios} is sorted to show the trend of rising PAH 15--20\mic\ to 11.3\mic\ ratio. For a
diffuse ISM mix of starlight-heated, neutral and ionized PAHs, 15--20\mic\ features are efficiently produced by
PAH particles in the 8--25\,\AA\ size range (300 \raisebox{0.5mm}{\scriptsize ${\lesssim}$} $ N_{\rm C}$
\raisebox{0.5mm}{\scriptsize ${\lesssim}$} $10^4$ carbon atoms). The 11.3\mic\ feature can be produced by
particles as large as about 20\,\AA\ or $ N_{\rm C}\approx4000$, but more efficiently by particles with radius
$a\,\raisebox{0.5mm}{\scriptsize ${\lesssim}$}\,10$\,\AA\ or $N_{\rm C}\,\raisebox{0.5mm}{\scriptsize
${\lesssim}$}\,500$ \citep{draine2006}. A direct, quantitative comparison of the observed PAH 15--20\mic\ to 11.3
and 6.2\mic, or the 16.4\mic\ to 11.3 and 6.2\mic\ ratios with model calculations is difficult for a number of
reasons. Foremost, there is a unique PAH size distribution, which differs from source to source. Secondly,
individual, discrete PAH features may be due to specific substructures in PAH molecules of a certain size
(cf.~interpretation of the 16.4\mic\ feature above). Finally, the specific local physical conditions affect the
charge state, the amount of dehydrogenation, and the heating process of the PAH molecules.

Nevertheless, it is illustrative to note some general trends when comparing the observed PAH emission ratios in
Table~\ref{tab:pahlineratios} with calculated PAH band radiation efficiencies as function of size for a diffuse
ISM mix of neutral/ionized PAHs \citep[cf.][]{draine2006}. We approximate the predicted, total emitted power in
the 15--20\mic\ region by scaling the calculated 17.4\mic\ band power with the total integrated absorption cross
section per C atom in the 15--20\mic\ range divided by the cross section for the 17.4\mic\ band. This gives
approximately a scale factor of 4, using the cross sections per C atom adopted by \citet{draine2006}. The lowest
observed 15--20/11.3, 15--20/6.2, 16.4/11.3, and 16.4/6.2\mic\ ratios in Table~\ref{tab:pahlineratios} correspond
then to the smallest PAHs considered in the calculations by \citet{draine2006}, which contain about 20 carbon
atoms. PAHs smaller than about 20--50 C atoms are photolytically unstable
\citep[cf.][]{allam1989,jochi1994,allai1996b,lepag2003}. Hence, the PAH size distribution of the sources near the
top of Table~\ref{tab:pahlineratios} is probably dominated by PAHs of that size. This is generally consistent with
earlier estimates of PAH sizes derived from the 3.3 and 3.4\mic\ C-H stretching modes
\citep[][Table~3]{allam1989}, and from profile fitting of the PAH 3.3-11.3\mic\ bands with calculated spectra from
a distribution of small PAHs \citep{verst2001,pech2002}.

For the supernova remnant N132D, we expect that small grains/PAHs are rapidly destroyed by thermal sputtering
(cf.~Sect.~\ref{sec;IRSgrainprocessing_discussion}), i.e.~relatively larger grains/PAHs tend to dominate the size
distribution, which is consistent with the high PAH 15--20\mic\ to 11.3\mic\ ratio and the absence of discrete
15--20\mic\ features. A PAH 15--20\mic\ to 11.3\mic\ ratio of $\sim$\,7 as observed in N132D corresponds to PAH
sizes of about 20\,\AA\ or 4000 C atoms, if we adopt the calculations by \citet{draine2006} as a tentative
guideline. This number is a rough estimate, since electron collisions rather than starlight dominate grain heating
in N132D, changing the amount and frequency of energy deposition into grains compared to the model calculations.
In addition, even large PAHs may be dehydrogenated in the hot plasma of N132D, which provides another rationale
for the weak 11.3\mic\ feature and the resulting high 15--20/11.3\mic\ ratio. As a consequence, the size estimate
from that ratio is probably an upper limit, since PAHs in N132D have most likely a higher degree of
dehydrogenation than in the diffuse ISM. A lower limit estimate of the PAH sizes in N132D from the 15--20/6.2\mic\
ratio is about 500 C atoms, which is similar to the sizes of the very small grains/large PAHs proposed as carriers
of the broad plateaus underlying the 6--14\mic\ PAH features \citep[e.g.][and
cf.~Sect.~\ref{sec:discussion_PAHlines} ]{allam1989,peete2002,rapac2005}.

The lifetime of 500--4000 C atom grains, about 0.001--0.002\mic\ in size, is 80--160\,yr according to
Eq.~\ref{eqn;thermsput}, which describes thermal sputtering of silicate/graphite grains in the forward shock of
N132D. We presume that large PAHs/amorphous carbon particles behave similarly to graphite in terms of stability
against thermal sputtering, which is a reasonable assumption since both consist mainly of aromatic carbon ring
systems. Thus, it is plausible for large PAHs to survive thermal sputtering in N132D long enough to maintain a
detectable population, particularly since both PAHs and grains will be replenished as the shock wave keeps
expanding into new ISM material.

\section{Summary}
We detected strong MIPS\,24\mic\ emission from swept-up, shock-heated dust grains in the young LMC supernova
remnant N132D (Sect.~\ref{sec;MIPSdiscussion}). In addition, we reported the first detection of PAH bands from a
SNR, with tentative PAH features between 6 and 9\mic, a weak 11.3\mic\ feature, and a prominent 15--20\mic\
emission hump. This adds to the growing body of sources showing 15--20\mic\ emission, which is interpreted as
\mbox{C-C-C} bending modes of large PAHs (Sect.~\ref{sec:discussion_PAHlines}).

Overall, our observations show that dust grains survive a strong supernova shock, which is consistent with our
estimates of the dust destruction rate by thermal sputtering (Sect.~\ref{sec;IRSgrainprocessing_discussion}). Our
results also support the proposed correlation of PAH size and the 15--20 to 11.3\mic\ emission ratio
\citep{vanke2000}. The ratio of 15--20\mic\ to 11.3\mic\ emission in N132D is notably higher compared to other
types of sources, which we interpret as rapid destruction of small PAHs/grains in the supernova blast wave via
thermal sputtering, and possible dehydrogenation of the remaining PAH population in the high temperature plasma.
We conclude that large PAHs/amorphous carbon particles with approximately 500--4000 C atoms survive the harsh
conditions in the forward shock of N132D long enough to be detected in our spectra
(Sect.~\ref{sec:previousobservation}). Future observations of supernova remnants of various ages located in
different environments will probe whether, and what fraction of dust grains and PAHs are ultimately able to
survive the aftermath of strong supernova shock waves.

\acknowledgements This work is based on observations made with the {\it Spitzer} Space Telescope, which is
operated by the Jet Propulsion Laboratory, California Institute of Technology under a contract with NASA. We thank
E.~Dwek, L.~Allamandola, and A.~Jones for fruitful discussions, and B.~Draine for providing a late draft of his
manuscript \citet{draine2006} prior to publication. We thank L.~Rudnick, T.~DeLaney, and U.~Hwang for their
collaboration on the {\it Spitzer} proposal leading to this work and their helpful comments. Support for this work
was provided by the NASA LTSA program, NRA-01-01-LTSA-013, and NASA funding through the {\it Spitzer} GO program.


\end{document}